\documentclass[prd,preprint,superscriptaddress,amsmath,amssymb,nofootinbib]{revtex4}
\usepackage{graphicx}
\usepackage{dcolumn}
\usepackage{bm}
\usepackage{amssymb}
\usepackage{amsmath}
\usepackage{epsfig}    
\usepackage{color}
\usepackage{slashed}
\usepackage{hhline}
\usepackage{bbm}

\def\be{\begin{equation}}
\def\ee{\end{equation}}
\newcommand{\bea}{\begin{eqnarray}}
\newcommand{\eea}{\end{eqnarray}}
\newcommand{\nn}{\nonumber}

\begin{document}

\title{A radiative seesaw model in a non-invertible selection rule with the assistance of a non-holomorphic modular $A_4$ symmetry }

\author{Shilpa Jangid }
\email{shilpajangid123@gmail.com}
\affiliation{Shiv Nadar IoE Deemed to be University, Gautam Buddha Nagar, Uttar Pradesh, 201314, India}

\author{ Hiroshi Okada}
\email{hiroshi3okada@htu.edu.cn}
\affiliation{Department of Physics, Henan Normal University, Xinxiang 453007, China}

\date{\today}

\begin{abstract}
We propose a two-loop neutrino mass model where fermionic and bosonic dark matter (DM) candidates are simultaneously connected to the neutrinos. But the fermionic DM candidate is favored compared to the bosonic one due to generating the fermionic DM mass at one-loop level.
In order to obtain our desired  Lagrangian and Higgs potential, we introduce a $Z_3$ gauging TY non-invertible fusion rule with the assistance of a non-holomorphic modular $A_4$ symmetry. The fusion rule forbids the mass of DM candidate at tree level but its mass is generated at one-loop level where the DM mass term dynamically violates the fusion rule.
After that, the neutrino mass matrix is induced at one-loop level where a remnant $Z_2$ symmetry is still remained.
The symmetry assures the stability of our DM candidate.
The non-holomorphic modular $A_4$ symmetry plays a role in forbidding the interactions between the SM particles and heavier fermions $X_R$ and an isospin singlet inert scalar boson $S_0$ that run in the DM mass loop, in addition to reduction our free parameters that leads to our predictions for lepton sector.
We perform $\chi^2$ numerical analysis for the lepton masses, mixing angles, and phases, and we show several predictions for NH and IH. Then, we demonstrate our lepton flavor violations, muon anomalous magnetic dipole moment, and the relic density of the DM candidate fixing the best fit point of the lepton sector.

 \end{abstract}
\maketitle

\section{Introduction}
Ma model~\cite{Ma:2006km} is one of the promising candidates to induce tiny neutrino masses by interacting with dark matter (DM) candidate at low energy scale. As a result, it appears on plenty of verifiable phenomenology such as lepton flavor violations (LFVs), and lepton anomalous magnetic dipole moment in addition to the DM phenomenology.
The model potentially has two kinds of dark matter (DM) candidates; a neutral component of isospin doublet inert boson ($\eta\equiv[\eta^+,(\eta_R+i\eta_I)/\sqrt2]^T$), and the lightest neutral Majorana fermion ($N_{R_1}$).
However, we have no ways theoretically to identify which one is an appropriate DM candidate.

In this paper,  we induce the masses of neutral fermions to be at one-loop level, applying the dynamical symmetry breaking of a non-invertible selection rule, which are recently applied to phenomenology~\cite{Jiang:2025psz, Kobayashi:2025rpx, Jangid:2025krp, Okada:2025kfm, Choi:2022jqy,Cordova:2022fhg,Cordova:2022ieu,Cordova:2024ypu,Kobayashi:2024cvp,Kobayashi:2024yqq,Kobayashi:2025znw,Suzuki:2025oov,Liang:2025dkm,Kobayashi:2025ldi,Kobayashi:2025lar,Kobayashi:2025cwx,Nomura:2025sod,Dong:2025jra,Nomura:2025yoa,Chen:2025awz, Kobayashi:2025thd, Suzuki:2025bxg}. 
Especially, $Z_3$ Tambara-Yamagami ($Z_3$-TY) selection rule possesses phenomenologically good nature whose algebra is as follows:
\begin{align}
&a\otimes a^*={\mathbbm 1},\ a \otimes a =a^*,\ a^*\otimes a^* = a,\ n\otimes n={\mathbbm 1}\oplus a\oplus a^*,\
a\otimes n =a^*\otimes n =n, 
\end{align}
where three elements $\{a,a^*,n\}$ are commutable each other.\\
Here, we denote the neutral fermions to be $N_R$ and their masses to be $M_N$. 
Since the mass of bosonic DM candidate ($m_0$) is generated at tree-level, we would insist on that the fermionic DM candidate is favored more than bosonic one in the scenario. $m_0^2(\equiv \frac{m_R^2+m_I^2}{2})$ is defined by the averaged mass of real part $m_R$ and imaginary part $m_I$ of neutral inert boson.
 In order to realize such one-loop fermionic masses, we need to introduce a new boson ($S_0\equiv (S_R+iS_I)/\sqrt2$) and fermions ($X_R$) that run inside the loop of $N_R$. But, these new particles have to interact with $N_R$ only. Otherwise, the model would be broken.
 To control these interactions, we need an additional symmetry such as a $Z_2$ discrete Abelian symmetry~\cite{Kajiyama:2013zla, Kajiyama:2013rla}, a gauged (global) hidden $U(1)$ symmetry~\cite{Nomura:2019vqc}, and so on. The simplest realization is to impose $Z_2$ odd for $S_0$ and $X_R$ and even for the other particles.~\footnote{Note that this $Z_2$ is different from the original $Z_2$ in Ma model.}
 Once the mass hierarchy $M_N \ll m_0$ can be realized, the neutrino mass matrix can approximately be simplified as follows~\cite{Baek:2015mna}:
 \begin{align}
 m_\nu \simeq
 \kappa \left(y_\eta M_N^* y^T_\eta\right),\label{eq:ma_type-II}
 \end{align}
  where dimensionless factor $\kappa$ consists of the loop factor and some parameters in this model, and 
 $y_\eta$ is Yukawa coupling of $\overline{L_L}\tilde \eta N_R$ with $\tilde \eta\equiv i\sigma_2 \eta^*$.
 Eq.~(\ref{eq:ma_type-II}) is known as the type-II structure. 
 Therefore, the neutrino mass structure is simpler than general form of the Ma model.
Thus, we introduce a non-holomorphic modular $A_4$ symmetry instead of $Z_2$ in order to reduce our free parameters and obtain some predictions for lepton sector~\footnote{Here, we show some models applying the non-holomorphic modular flavor symmetries~\cite{Qu:2024rns, Ding:2024inn, Li:2024svh, Nomura:2024atp, Nomura:2024vzw, Nomura:2024nwh,Qu:2025ddz, Okada:2025jjo, Kobayashi:2025hnc, Loualidi:2025tgw, Nomura:2025ovm, Abbas:2025nlv, Nomura:2025raf}.}. Then, we will discuss the other phenomenologies such as LFVs, muon anomalous magnetic dipole moment, and the nature of DM candidate and show numerical results.

This paper is organized as follows. In Section~II, we present our model setup.
In Section III, we perform numerical analysis to search for allowed parameter region and predicted flavor observables. 
Finally we devote Section~IV to the summary and conclusion.

\begin{table}[!ht]
\begin{center}
\begin{tabular}{|c||c|c|c|c|c||c|c|}\hline\hline
& \multicolumn{5}{c||}{Ma\ model} & \multicolumn{2}{c|}{Hidden sector} \\ \hline
Fields & $\overline{L_L}$ & $\ell_R$ & $N_R$ & ~$H$~ & ~$\eta^*$~  & ~~$X_R$~~ & $S_0$ \\ \hline
$SU(2)_L$ & $\mathbbm{2}$ & $\mathbbm{1}$ & $\mathbbm{1}$ & $\mathbbm{2}$   & $\mathbbm{2}$ & $\mathbbm{1}$ & $\mathbbm{1}$ \\ \hline
$U(1)_Y$ & $\frac{1}{2}$ & $-1$ & $0$ & $ \frac{1}{2}$& $-\frac{1}{2}$  & $0$ & $0$ \\ \hline
$\mathcal{Z}_3$-${\rm TY}$ & $n$ & $n$ & $a$ & $\mathbbm{1}$  & $n$ & $n$ & $n$ \\ \hline
$({A}_4)_{-k_I}$ & $\{\overline{\bf1}\}_{-1}$ & $\{{\bf 1}\}_{1}$ & $\{{\bf1}\}_{-1}$ & ${\bf1}_{0}$  
& ${\bf1}_{2}$ & ${\bf3}_{0}$ & ${\bf1}_{1}$ \\ \hline 
\end{tabular}
\caption{Field contents and their charge assignments under the $SU(2)_L\otimes U(1)_Y\otimes \mathcal{Z}_3$-${\rm TY}\otimes A_4$}, where $-k_I$ in the lower index of $A_4$ is the number of modular weight. \label{tab:fieldcontent}
\end{center}
\end{table}

\section{Model setup}
\label{sec:II}
Here, we review our model.
In addition to the field contents of Ma model, we introduce three neutral fermions $X_R$ and an inert singlet boson $S_0$.
Both these particles have $\sigma$ under $Z_3$-TY. ${\bf 3}_0$ and ${\bf 1}_1$ are respectively imposed for $X_R$ and $S_0$ under modular $A_4$ symmetry where the lower index of each representation represents the modular weight. 
The field contents and their charge assignments of Ma model as well as new particles are summarized in Table \ref{tab:fieldcontent},
where $\{{\bf1}\}\equiv \{{\bf1},{\bf1'},{\bf1''}\}$ and $\{\overline{\bf1}\}\equiv \{{\bf1},{\bf1''},{\bf1'}\}$.
At first let us discuss the allowed renormalizable terms without the modular $A_4$ symmetry,  which are found as follows:
\begin{align}
&y_{\ell_i}  \overline{L_{L_i}} H \ell_{R_i} + y_{\eta_{ib}} \overline{L_{L_i}} \tilde \eta N_{R_a} +\lambda_5 (H^\dag\eta)^2
+ Y_{a\alpha} \overline{N^C_{R_a}} X_{R_\alpha} S_0 + M_{X_{\alpha\beta}} \overline{X^C_{R_\alpha}} X_{R_\beta} +\mu_S^2 S_0^2  + {\rm h.c.}  \label{eq:1} \\
&
+g_{i\alpha}  \overline{L_{L_i}} \tilde \eta X_{R_\alpha}  + \mu \eta^\dag H S_0 + {\rm h.c.}, \label{eq:2}
\end{align}
where we abbreviate trivial terms in the Higgs potential. 
As can be seen in Eq.~(\ref{eq:1}), the mass of $N_R$ at tree level is forbidden by the $Z_3$-TY selection rule, but allowed at one-loop level via
$Y_{a\alpha} \overline{N^C_{R_a}} X_{R_\alpha} S_0 + M_{X_{\alpha\beta}} \overline{X^C_{R_\alpha}} X_{R_\beta} +\mu_S^2 S_0^2$
where the $Z_3$-TY is dynamically broken. Needless to say, $ \overline{L_{L_i}} \tilde H N_{R_a}$ is also forbidden by this selection rule.
However, the $Z_3$-TY symmetry cannot forbid terms $g_{i\alpha}  \overline{L_{L_i}} \tilde \eta X_{R_\alpha}  + \mu \eta^\dag H S_0$ in Eq.~(\ref{eq:2}). $g_{i\alpha}  \overline{L_{L_i}} \tilde \eta X_{R_\alpha}$ leads to the standard Ma model together with $M_{X_{\alpha\beta}} \overline{X^C_{R_\alpha}} X_{R_\beta}$ and $\lambda_5 (H^\dag\eta)^2$.
And $ \mu \eta^\dag H S_0$ gives us additional one-loop neutral fermion mass matrices by mixing between $\eta$ and $S_0$.
Therefore, we need to vanish these terms in Eq.~(\ref{eq:2}). 
Taking the modular $A_4$ symmetry into account, these terms are forbidden;
especially, both terms are forbidden by odd modular weights.
It goes without saying that all needed terms in Eq.~(\ref{eq:1}) are allowed by the modular $A_4$ flavor symmetry.
\footnote{On the contrary, only with the modular $A_4$ symmetry, $ \overline{L_{L}} \tilde H N_{R}$ and $ \overline{N^C_{R}} N_{R}$ cannot be forbidden where these two terms lead us to the boring canonical seeesaw.}

\subsection{The mass matrices for $X_R$ and $N_R$}
The mass matrix of $N_R$ is induced via the following terms:
\begin{align}
& Y_{a\alpha} \overline{N^C_{R_a}} X_{R_\alpha} S_0 + M_{X_{\alpha\beta}} \overline{X^C_{R_\alpha}} X_{R_\beta} +\mu_S^2 S_0^2  + {\rm h.c.}  \label{eq:NR-mass} 
\end{align}
However since the mass matrix of $X_R$ is not diagonal due to the modular $A_4$ flavor symmetry, we start diagonalizing the mass matrix of $X_R$. $M_X$ is found as
\begin{align}
M_X = M_3 
\left[\begin{array}{ccc}
\tilde{M}_0+ 2y_{1}^{(0)}  & -y_{3}^{(0)}  & -y_{2}^{(0)}  \\ 
-y_{3}^{(0)} & 2y_{2}^{(0)} & \tilde{M}_0-y_{1}^{(0)} \\ 
-y_{2}^{(0)} & \tilde{M}_0-y_{1}^{(0)} &2 y_{3}^{(0)} \\ 
\end{array}\right],
\end{align}
where $Y^{(0)}_{3}\equiv [y_1^{(0)},y_2^{(0)},y_3^{(0)}]^T$ is $A_4$ triplet with zero modular weight, and $\tilde M_0$ is dimensionless free parameter. $M_3$ is supposed to be a real mass parameter.
Then, $M_X$ is diagonalized by a unitary matrix $U_X$; $\tilde D_X\equiv D_X/M_3=U_X^T M_X U_X$. Therefore, the flavor eigenstate of $X_R$ is rewritten in terms of mass eigenstate of $\psi_R$ as $X_R = U_X \psi_R$.

Now that $X_R$ is diagonalized,  we move on to evaluate the mass matrix of $N_R$.
$M_N$ at one-loop level is induced via the following terms:
\begin{align}
\left(\frac{Y_{a\alpha}U_{X_{\alpha\beta}}}{\sqrt2}\right)
 \overline{N^C_{R_a}} X_{R_\beta} (S_R+ i S_I) + {\rm h.c.}, 
\end{align}
where $Y$ is explicitly given by
\begin{align}
Y=
\left[\begin{array}{ccc}
\alpha_N  &0 & 0 \\ 
0& \beta_N & 0\\ 
0 & 0 & \gamma_N \\ 
\end{array}\right]
\left[\begin{array}{ccc}
y_{1}^{(0)}  & y_{3}^{(0)}  & y_{2}^{(0)}  \\ 
y_{3}^{(0)} & y_{2}^{(0)} & y_{1}^{(0)} \\ 
y_{2}^{(0)} & y_{1}^{(0)} & y_{3}^{(0)} \\ 
\end{array}\right],
\end{align}
where $\alpha_N,\ \beta_N,\ \gamma_N$ are dimensionless complex free parameters.  
The explicit form of $M_N$ is found as
\begin{align}
M_{N_{ab}} =\frac{M_3}{2(4\pi)^2}
(Y U_X)_{a\beta} \tilde D_{X_\beta} (Y U_X)^T_{\beta b}
\left[
\frac{\tilde m^2_R}{\tilde m^2_R - \tilde D^2_{X_{\beta}}}\ln\left(\frac{\tilde m^2_R}{ \tilde D^2_{X_{\beta}}}\right)
-
\frac{\tilde m^2_I}{\tilde m^2_I - \tilde D^2_{X_{\beta}}}\ln\left(\frac{\tilde m^2_I}{ \tilde D^2_{X_{\beta}}}\right)
\right].
\end{align}
Here, $\tilde m_R\equiv m_R/M_3$ and $\tilde m_I\equiv m_I/M_3$ are respectively the mass eigenvalues of $S_R$ and $S_I$. 
We redefine the following mass combinations; $\tilde m_S^2\equiv (\tilde m_R^2+\tilde m_I^2)/2$ and 
 $\delta\tilde m^2\equiv (\tilde m_R^2 - \tilde m_I^2)/2$, assuming $\delta\tilde m^2 \ll \tilde m_S^2$.~\footnote{Even though it is nothing but assumption by hand, it does not affect the structure of mass matrix of $N_R$.}
Then, we simplify $M_N$ as follows:
\begin{align}
M_{N_{ab}} &\simeq - M_3  \frac{\delta\tilde m^2}{(4\pi)^2}
\frac{(Y U_X)_{a\beta}  (Y U_X)^T_{\beta b}}{\tilde D_{X_\beta}}
\frac1{(1-r_\beta)^2}
\left[{1- r_\beta +\ln(r_\beta)}\right]\\
&-M_3 \tilde M_{N_{ab}} ,
\end{align}
where $r_\beta \equiv \tilde m^2_S/\tilde D^2_{X_\beta}$.
Similar to the case of $X_R$, $N_R$ is diagonalized by the unitary matrix $U_N$, and we find the following relations $\tilde D_N(\equiv D_N/M_3) = U_N^T \tilde M_N U_N$ and $N_R= U_N\Psi_R$.

\subsection{Active neutrino mass matrix}
The tiny neutrino mass matrix is given by the following terms:
\begin{align}
&a_\eta \frac{(y_\eta U_N)_{ia}}{\sqrt2}\overline{\nu_{L_i}}\Psi_{R_a}(\eta_R - i \eta_I) +{\rm h.c.}, \\
& y_\eta={\rm diag}(1,\beta_\eta,\gamma_\eta).
\end{align}
Then, under $M_N\ll m_0$, the neutrino mass matrix is found to be 
\begin{align}
m_{\nu_{ij}} &\approx - M_3 \frac{\lambda_5 (a_\eta v_H)^2}{8\pi^2 m^2_0} (y_\eta U_N)_{ia} \tilde D^*_{N_a} (y_\eta U_N)_{aj}\\
& =  - \frac{\lambda_5 (a_\eta v_H)^2}{(4\pi)^2 M_3 \tilde m^2_0} \left(y_{\eta_i} \tilde M^*_{N_{ij}} y_{\eta_j}\right)\\
&\equiv \kappa_\nu \tilde m_{\nu_{ij}},
\end{align}
where $m_0\equiv M_3 \tilde m_0$ and
\begin{align}
\kappa_\nu\equiv  - \frac{\lambda_5 (a_\eta v_H)^2}{(4\pi)^2 M_3 \tilde m^2_0} .\label{eq:kappa}
\end{align}
Then, the active neutrino mass matrix is diagonalized by a Unitary matrix $U_\nu$ as
$
(\tilde m_1,\tilde m_2,\tilde m_3)=U^T_\nu \tilde m_\nu U_\nu$.
$\kappa_\nu$ is determined by atmospheric mass square difference and mass eigen-values as follows: 
\begin{align}
({\rm NH}):\ \kappa_{\nu}^2 = \frac{\Delta m^2_{\rm atm}}{\tilde m^2_3-\tilde m^2_1}, \label{eq:nh} \\
({\rm IH}):\ \kappa_{\nu}^2 = \frac{\Delta m^2_{\rm atm}}{\tilde m^2_2-\tilde m^2_3},\label{eq:ih}
\end{align}
where NH is normal hierarchy and IH represents inverted hierarchy.
The solar mass square difference is then determined by
 \begin{align}
 \Delta m^2_{\rm sol} =  \kappa_\nu^2 (\tilde m^2_2-\tilde m^2_1).
\end{align}
Hereafter, we use a single notation of $\kappa_\nu$, but it implicitly depends on hierarchy as can be seen in Eq.~(\ref{eq:nh}) and Eq.(\ref{eq:ih}).
In our numerical analysis, $\Delta m^2_{\rm atm}$ is given as the input parameter from experimental result.
Then, $\Delta m^2_{\rm sol}$ is evaluated as our prediction and compared with the experimental result.
Sum of the neutrino masses $\sum_i m_i$ is given by
\begin{align}
\sum_i m_i = \kappa_\nu(\tilde m_1 + \tilde m_2 + \tilde m_3).
\end{align} 
$\sum_i m_i $ has upper bound on 120 meV~\cite{Vagnozzi:2017ovm, Planck:2018vyg} from
the minimal cosmological model
$\Lambda$CDM,
and recently has $\sum m_{i}\le$ 72 meV~\cite{DESI:2024mwx}  from the combination of DESI and CMB. 
Due to the modular flavor symmetry with an appropriate assignment, the charged-lepton mass matrix is diagonal basis.
Thus, $U_{\rm PMNS}= U_\nu$ where $U_{\rm PMNS}$ has observables and we adopt the standard parametrization as follows~\footnote{Without loss of generality, any unitary matrix for lepton sector can be written in terms of three mixing angles and three phases via phase redefinitions for fields.}:
\begin{align}
 \left[\begin{array}{ccc}
1 & 0  &0 \\ 
0 & c_{23} & s_{23} \\ 
0& -s_{23} & c_{23} \\ 
\end{array}\right]
 \left[\begin{array}{ccc}
c_{13} & 0  & s_{13}e^{-i\delta_{\rm CP}} \\ 
0 & 1 & 0 \\ 
-s_{13}e^{-i\delta_{\rm CP}} & 0 & c_{13} \\ 
\end{array}\right]
\left[\begin{array}{ccc}
c_{12} & s_{12}  &0 \\ 
-s_{12} & c_{12} & 0 \\ 
0& 0 & 1 \\ 
\end{array}\right]
\left[\begin{array}{ccc}
1 & 0  &0 \\ 
0 & e^{i\alpha/2} & 0 \\ 
0& 0 & e^{i\beta/2} \\ 
\end{array}\right].
 \end{align}
 Then, the each of three mixing angles is given in terms of components of $U_{\rm PMNS}$ as follows:
 \begin{align}
 s_{13} = |(U_{\rm PMNS})_{13}|,\quad
 s_{12} = \frac{|(U_{\rm PMNS})_{12}|}{c_{13}},
\quad
 s_{23} = \frac{|(U_{\rm PMNS})_{23}|}{c_{13}},
 \end{align}
 where $s_{13}(c_{13}),s_{12},s_{23}$ are respectively abbreviated symbols for $\sin\theta_{13}(\cos\theta_{13}),\sin\theta_{12},\sin\theta_{23}$.
 Dirac CP phase $\delta_{\rm CP}$ and Majorana phases $\alpha,\ \beta$ are found as 
 \begin{align}
&\delta_{\rm CP} = {\rm Arg}[(U_{\rm PMNS})_{23}] - {\rm Arg}[(U_{\rm PMNS})_{13}],\\
& \alpha = 2{\rm Arg}[(U_{\rm PMNS})_{12} ],
\quad
 \beta = 2 {\rm Arg}[(U_{\rm PMNS})_{23} ].
 \end{align}
 The neutrinos double beta decay, which is denoted by $m_{ee}$, is found by
   \begin{align}
m_{ee}
=
\kappa_\nu
\left|
\tilde m_1 c^2_{12} c^2_{13} + \tilde m_2 s^2_{12}c^2_{13} e^{i\alpha} + \tilde m_3 s^2_{13} e^{i(\beta-2\delta_{\rm CP})}
\right| .
\end{align}
 $\langle m_{ee}\rangle$ has  upper bounds $(28–122 )$ meV at 90 \% confidence level from
a current KamLAND-Zen data~\cite{KamLAND-Zen:2024eml}.

\subsection{L FVs and muon $g-2$}
\label{sec:leptonpheno}
New contributions for LFVs arise from $y_\eta$. 
First of all, let us consider
the processes $\ell_\alpha \to \ell_\beta \gamma$ at one-loop
level.
Inserting Eq.~(\ref{eq:kappa}), our formulae for the branching ratio are written as
\begin{align}
&{\rm BR}(\ell_i \to \ell_j \gamma)
=
\frac{48\pi^3 \alpha_{\rm em}}{(\lambda_5 M_3 v_H^2 {\rm G_F} )^2 }\,
\tilde m^4_0 \kappa_\nu^2  \\  
&\hspace{0.5cm}\times 
C_{ij} 
\left(
\left| (y_\eta U_N)_{ja} F_{lfv}(\tilde D_{N_a},\tilde m_0)(y_\eta U_N)^\dag_{ai}\right|^2
+
\frac{m^2_{\ell_j}}{m^2_{\ell_i}}
\left| (y_\eta U_N)_{ia} F_{lfv}(\tilde D_{N_a},\tilde m_0)(y_\eta U_N)^\dag_{aj}\right|^2
\right),\nn\\
& F_{lfv}(m_1,m_2)
\simeq 
\frac{2 m^6_1+3m_1^4m_2^2-6m_1^2m_2^4+m_2^6+6m_1^4m_2^2\ln\left(\frac{m_2^2}{m_1^2}\right)}{12(m_1^2-m_2^2)^4},
\end{align}
where $\alpha_{\rm em}\approx1/137$ is the fine-structure constant,
$C_{ij} \approx(1,0.1784, 0.1736)$ for ($(i,j)=(\mu,e),(\tau,e),(\tau,\mu)$), 
${\rm G_F}\approx1.17\times 10^{-5}$ GeV$^{-2}$ is the Fermi constant.

 \begin{table}[t]
\begin{tabular}{c|c|c|c} \hline
Process & $(i,j)$ & Experimental bounds & References \\ \hline
$\mu^{-} \to e^{-} \gamma$ & $(\mu,e)$ &
	${BR}(\mu \to e\gamma) < 4.2 \times 10^{-13}$  ($90\%$ CL)& \cite{MEG:2016leq} \\
$\tau^{-} \to e^{-} \gamma$ & $(\tau,e)$ &
	${BR}(\tau \to e\gamma) < 3.3 \times 10^{-8}$  ($90\%$ CL)& \cite{BaBar:2009hkt} \\
$\tau^{-} \to \mu^{-} \gamma$ & $(\tau,\mu)$ &
	${BR}(\tau \to \mu\gamma) < 4.4 \times 10^{-8}$ ($90\%$ CL) & \cite{BaBar:2009hkt}   \\ \hline
$\Delta a_\mu$ & $(\mu,\mu)$ &
	$ (38\pm63) \times 10^{-11}$ ($1\sigma$) & \cite{Aliberti:2025beg}   \\ \hline
\end{tabular}
\caption{Summary for the experimental bounds of the LFV processes 
$\ell_\alpha \to \ell_\beta \gamma$ and muon $g-2$.}
\label{tab:Cif}
\end{table}

Similar to the computations for LFVs, our formula for the lepton $g-2$ can be written by
\begin{align}
\Delta a_\ell
\approx 2
\frac{\tilde m_0^2}{\lambda_5} \frac{m^2_{\ell_\ell}\kappa_\nu}{v_H^2 M_3} 
 (y_\eta U_N)_{\ell a} F_{lfv}(\tilde D_{N_a},\tilde m_0)(y_\eta U_N)^\dag_{a\ell}
.\label{damu}
\end{align}

\section{Dark matter candidate}
\label{sec:III}
In our model, $\chi_R \equiv N_{R_1}$ is a good DM candidate of Majorana fermion, where we redefine its mass to be $m_\chi(\equiv D_{N_1})$
or dimensionless form $\tilde m_\chi\equiv m_\chi/M_3$.
Since our DM does not couple to quark sector at tree-level, we simply neglect the constraints from direct detection searches.
%
To evaluate the observed relic density,
we impose the following condition to analyze it simpler;
$1.2 m_\chi\lesssim D_{N_2}\le D_{N_3}$, in order to evade an effect of co-annihilation interactions.
Under the condition, the dominant contribution to the relic density arises from $y_\eta$
and the non-relativistic cross section is expanded by relative velocity $v_{\rm rel}$; $(\sigma v_{\rm rel})\approx a_{\rm eff}
 + b_{\rm eff} v^2_{\rm rel}+ {\cal O}(v^4_{\rm rel})$.
Due to the Majorana nature, we have p-wave dominant contribution under the vanishing charged-lepton masses and its form is given by
\begin{align}
 b_{\rm eff} 
 \approx
\frac{(4\pi)^3}{3}
\left(\frac{\tilde m_0^2 \kappa_\nu}{\lambda_5 v^2_H \tilde m_\chi}\right)^2
R^2 (1-2R+2R^2) |y_\eta U_N|^2_{11} , 
\end{align}
where $R\equiv m_\chi^2/(m^2_0+m^2_\chi)$.
Then, the  relic density is given by
\begin{align}
\Omega h^2\approx 
\frac{1.07\times10^9 }{\rm GeV}
\frac{x_f^2}{3 \sqrt{g^*} M_P b_{\rm eff} 
},
\end{align}
where $g^*\approx100$, $M_P\approx 1.22\times 10^{19}{\rm GeV}$, $x_f\approx20$.
The experimental result at Planck tells us, $\Omega h^2=0.1196\pm0.0031$ at 1$\sigma$~\cite{Planck:2013pxb}.
In our numerical analysis below, we impose the above result within 2$\sigma$.

 \section{Numerical results}
 \label{sect.3}
 Here, we perform our numerical $\chi^2$ analysis adopting Nufit 6.0~\cite{Esteban:2024eli} as five reliable experimental observables; three mixings ($s_{12},\ s_{13},\ s_{23}$), two mass squared differences ($\Delta m^2_{\mathrm{atm}},\ \Delta m^2_{\mathrm{sol}}$). We treat with three phases; ($\delta_{\rm CP},\ \alpha,\ \beta$), as output predictions.
Then, we randomly select our input parameters within the following ranges 
\begin{align}
& (|\alpha_N|,|\beta_N|,|\gamma_N|,|\beta_\eta|,|\gamma_\eta|) \in [10^{-3},10],\
(\tilde M_0,\ \tilde m_S) \in [10^{-5},1],\
 \delta \tilde m \in [10^{-2},1], \label{eq:free_1} \\
& \lambda_5 \in [10^{-10},1],\
(M_3,\ m_0) \in [10^{2},10^{10}] \ {\rm GeV},\label{eq:free_2}
\end{align}
where we work on fundamental region of $\tau$. 
$\lambda_5,\ M_3,\ m_0$ are applied for LFVs, muon $g-2$, and DM relic density,  under the best fit value (BF) of our analysis in cases of NH and IH.

\subsection{NH}
\begin{figure}[tb]
\begin{center}
\includegraphics[width=50.0mm]{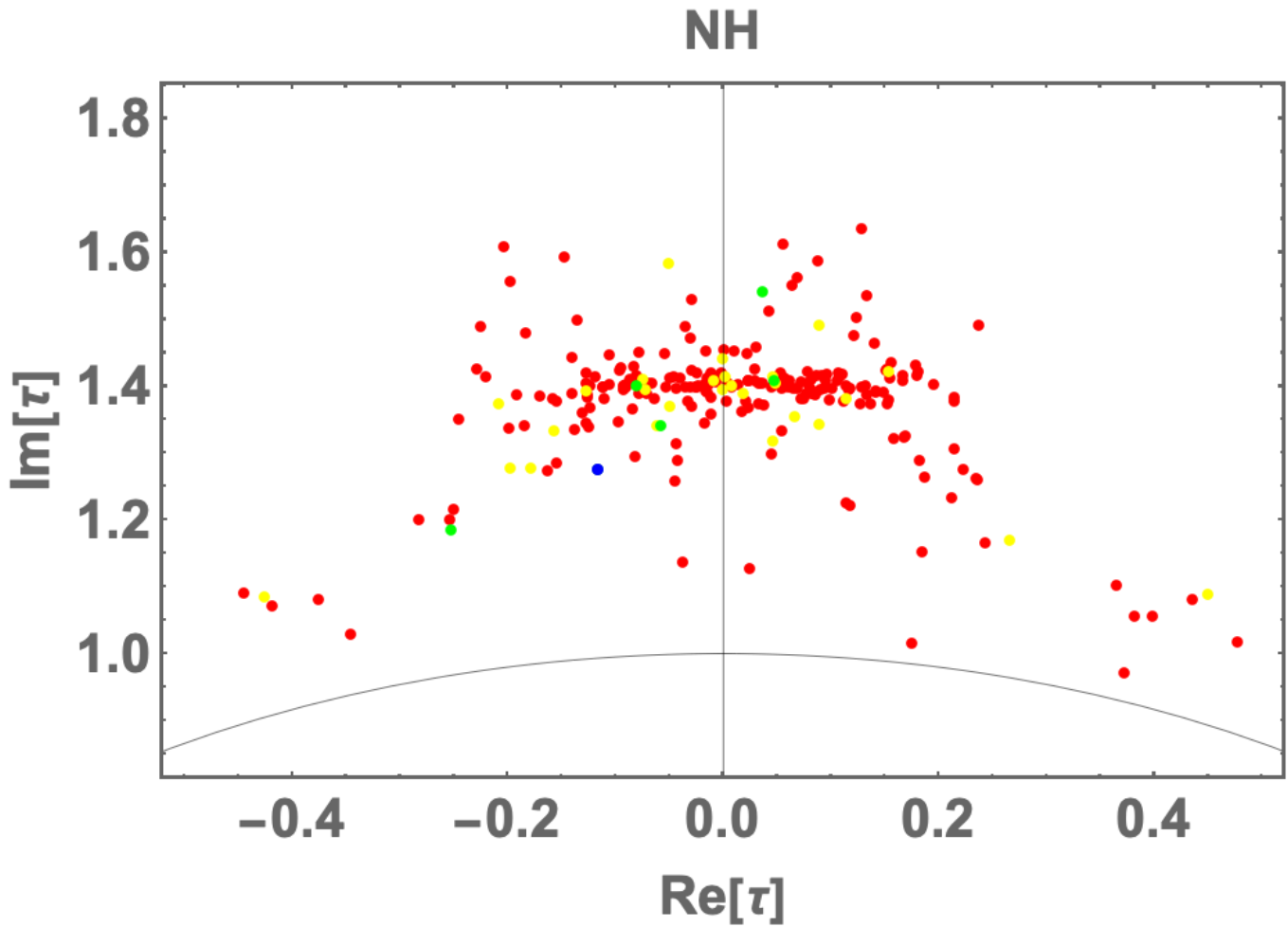} \quad
\caption{Allowed region for Im[$\tau$] in terms of Re[$\tau$] within the fundamental region through the numerical $\chi^2$ analysis.
Here, the red points are within 5$\sigma$, yellow ones within 3$\sigma$, green ones within 2$\sigma$, and, blue ones within 1$\sigma$.}
  \label{fig:tau_nh}
\end{center}\end{figure}

In Fig.~\ref{fig:tau_nh}, we show allowed region of $\tau$,
where the red points are within 5$\sigma$, yellow ones within 3$\sigma$, green ones within 2$\sigma$, and, blue ones within 1$\sigma$. The allowed region of Re[$\tau$] runs whole the fundamental region, on the other hand
the allowed one of Im[$\tau$] is allowed up to 1.65.

\begin{figure}[tb]
\begin{center}
\includegraphics[width=50.0mm]{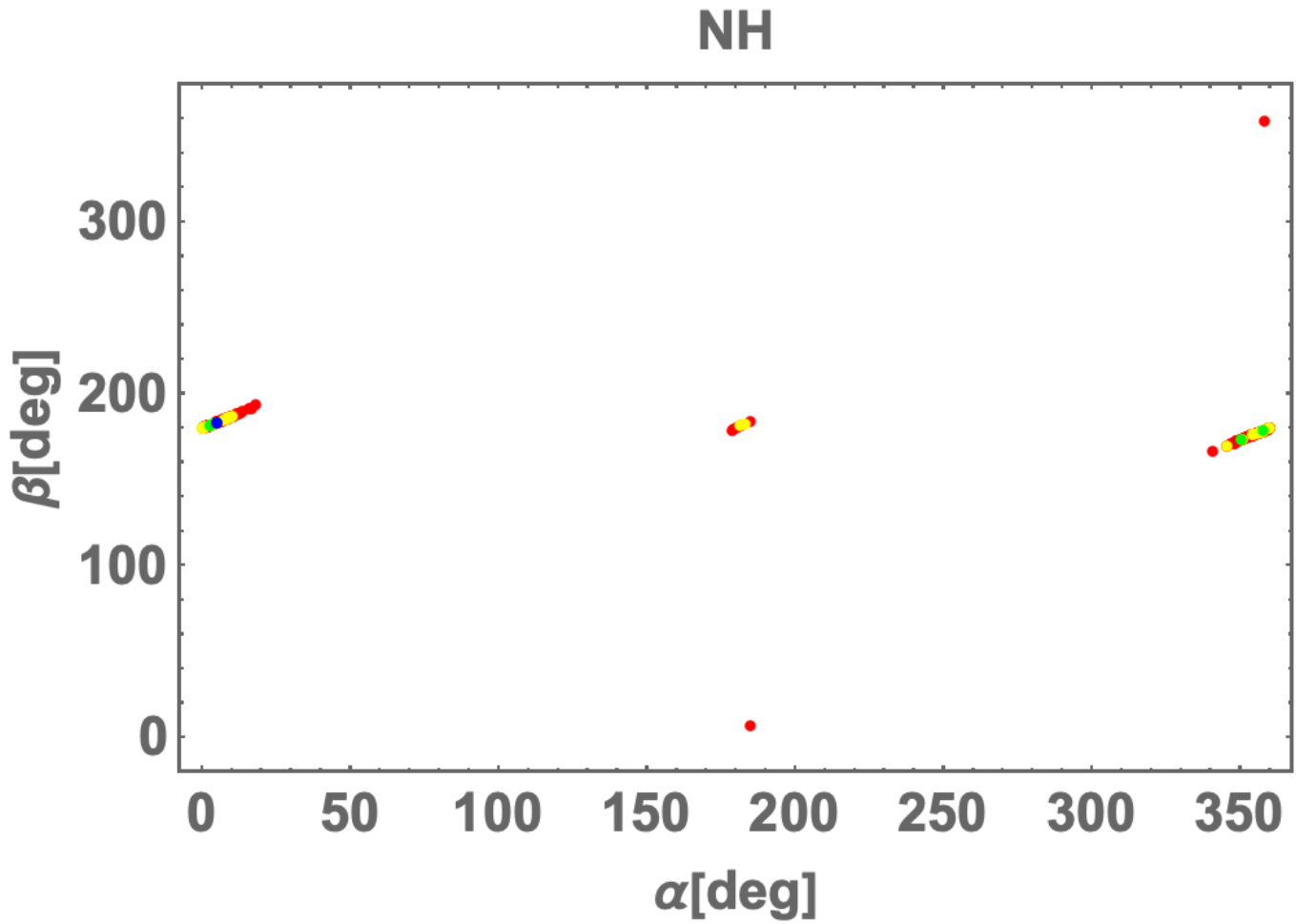} \quad
\includegraphics[width=50.0mm]{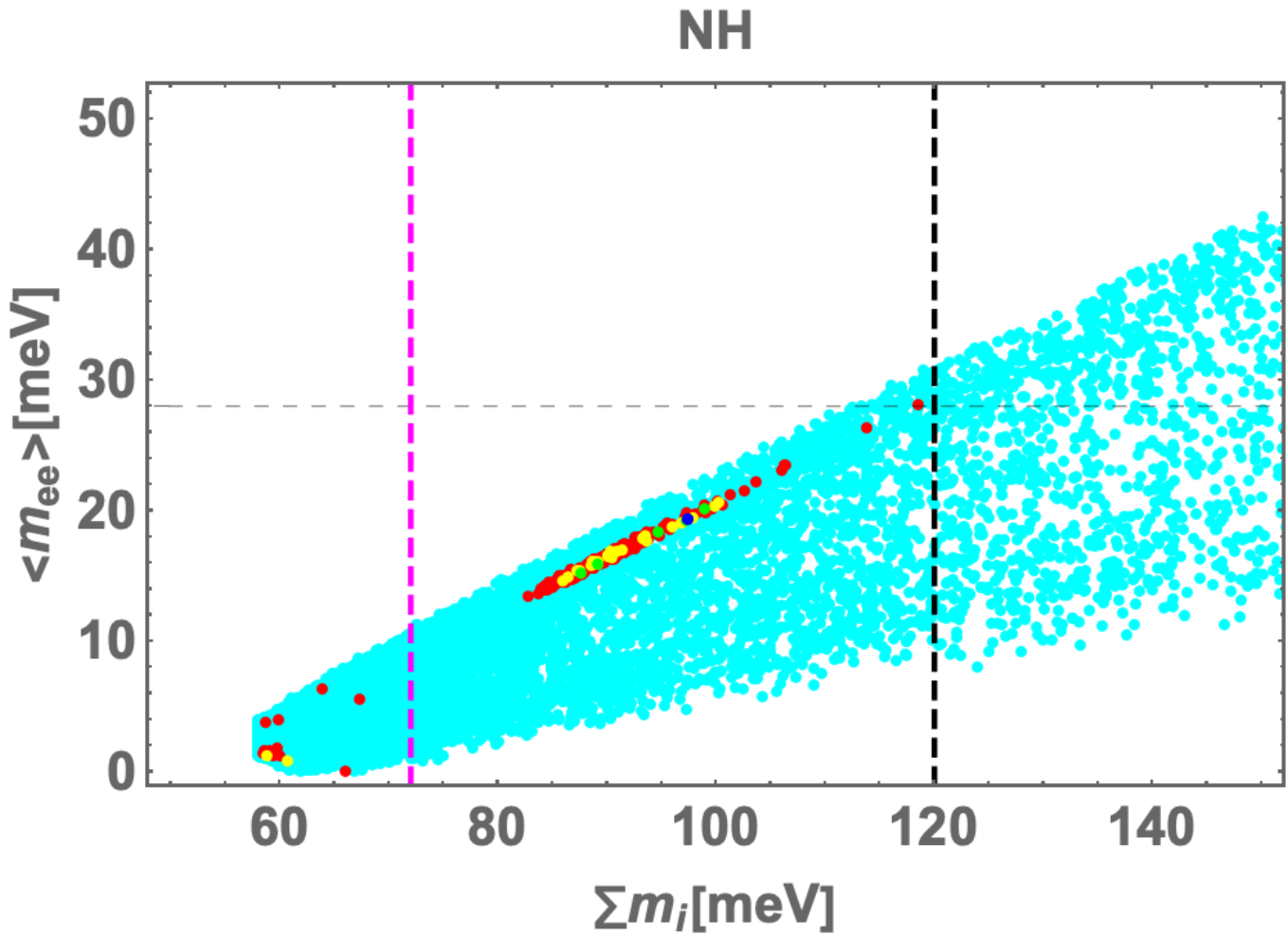} \quad
\includegraphics[width=50.0mm]{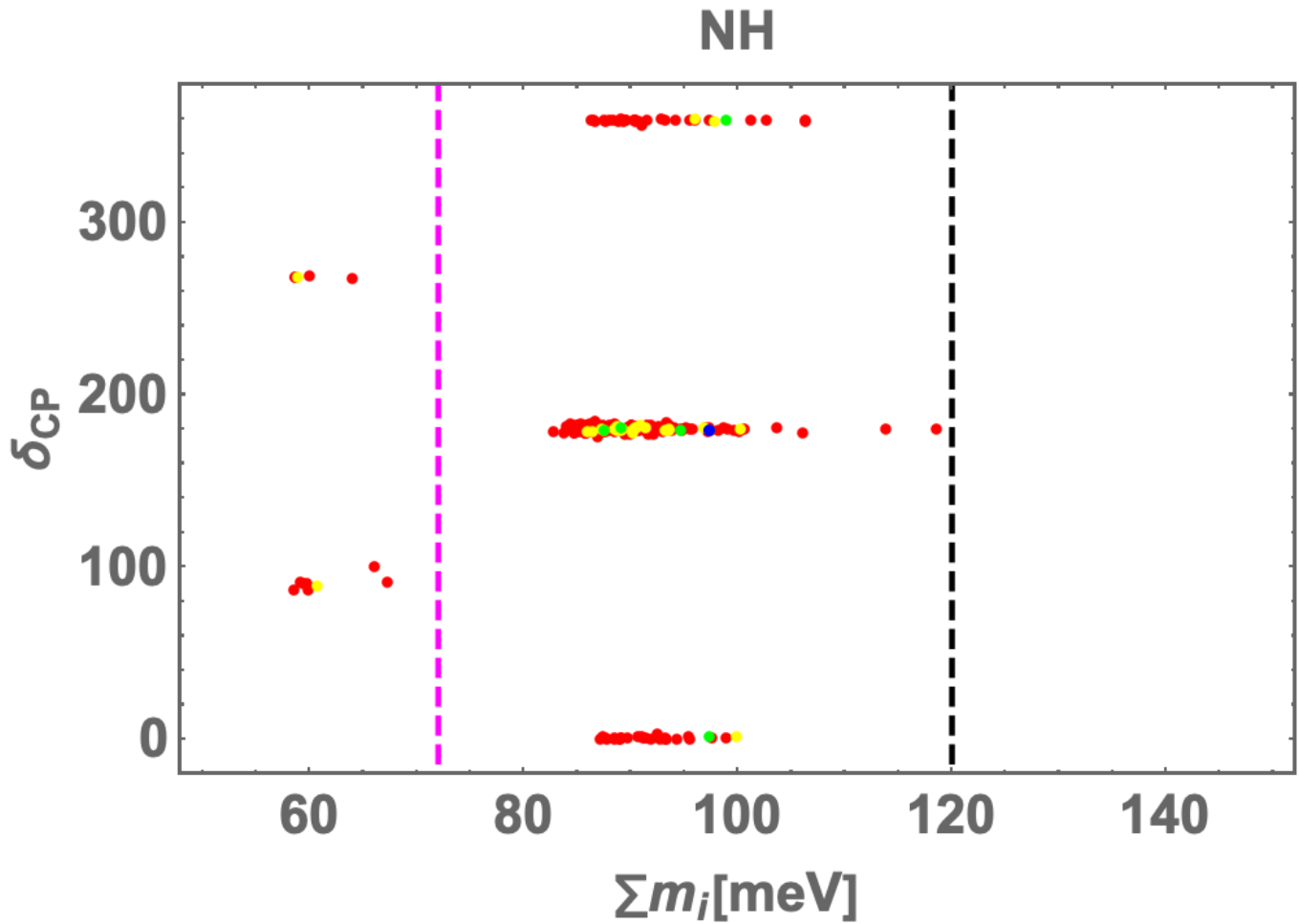} 
\caption{Allowed regions for Majorana phases (left), $\langle m_{ee}\rangle$ (center), and $\delta_{\rm CP}$ (right) in term $\sum m_i$ meV, where all the color legends are the same as the ones of Fig.~\ref{fig:tau_nh}. In the center, the cyan region is experimentally allowed, the magenta and black dotted vertical lines (in the center and right) are respectively the upper bounds of 72 meV (DESI and CMB) and 120 meV (the minimal cosmological model), the horizontal dotted line at 28 meV (in the center) represents the lower upper bound on KamLAND-Zen data. }
  \label{fig:nh2}
\end{center}\end{figure}
In Fig.~\ref{fig:nh2}, we demonstrate the allowed regions for Majorana phases (left), $\langle m_{ee}\rangle$ (center), and $\delta_{\rm CP}$ (right) in term $\sum m_i$ meV, where all the color legends are the same as the ones of Fig.~\ref{fig:tau_nh}. In the center, the cyan region is experimentally allowed, the magenta and black dotted vertical lines (in the center and right) are respectively the upper bounds of 72 meV (DESI and CMB) and 120 meV (the minimal cosmological model), the horizontal dotted line at 28 meV (in the center) represents the lower upper bound on KamLAND-Zen data. 
We have five localized islands in allowed points of Majorana phases and two of Majorana phases are at nearby $0^\circ$ or $180^\circ$. The sum of neutrino masses reaches up to 120 meV and the neutrinoless double beta decay allows up to 28 meV.
The allowed regions of Dirac CP phase have four specific; $0^\circ$, $100^\circ$, $180^\circ$, and $270^\circ$. 
%
\begin{table}[tb]
    \setlength\tabcolsep{0.2cm}
    \begin{tabular}{c|c||c|c||c|c}
\hline
        parameter    &  BF & parameter & BF & parameter & BF \\ \hline \hline
          $s_{12}$ &  $0.549$ & $s_{23}$ & 0.642 &$s_{13}$ &0.148 \\ \hline
       $\Delta m^2_{\rm sol}$  & $7.43\times10^{-23}$ GeV$^2$ 
       &  $\Delta m^2_{\rm atm}$ & $2.60\times10^{-21}$ GeV$^2$  &
 $\kappa_\nu$ &   $2.53\times10^{-7}$ \\ \hline
 $\langle m_{ee}\rangle$ & 15.6 meV  &
$\sum m_i $ & 88.8 meV & $m_1$ &   16.6 meV \\ \hline
$\tau$ & $0.0947+1.41 i$ &$\beta_\eta$ & $0.0247 - 0.216 i$  & $\gamma_\eta$ & $0.103 - 0.0189 i$ \\ \hline
$\alpha_N$ & $-0.00287 - 0.00614 i$ 
&$\beta_N$ & $-0.0173 - 0.0159 i$  
& $\gamma_N$ & $-0.0697 - 0.0102i$  \\
\hline
$\tilde M_0$ & $0.0202$ 
&$\tilde m_S$ & $0.000105$  
& $\delta\tilde m^2$ & $0.303$  \\
\hline
    \end{tabular}
    \caption{\label{tab:BF_nh}%
      Best-fit (BF) experimental values in the NH case corresponding to $\Delta\chi^2_{\rm min} \approx 0.0249$.}
\end{table}

In order to analyze LFVs, muon $g-2$, and fermionic DM candidate in our model,
we at first fix our benchmark point (BP) so as to be the best fit (BF) of $\chi^2$.
In Table \ref{tab:BF_nh}, we show BP that is $\Delta\chi^2_{\rm min} \approx 0.0249$.
Then, we randomly select our three free parameters $\lambda_5,\ M_3,\ m_0$ in Eq.~(\ref{eq:free_2}).
%
\begin{figure}[tb]
\begin{center}
\includegraphics[width=50.0mm]{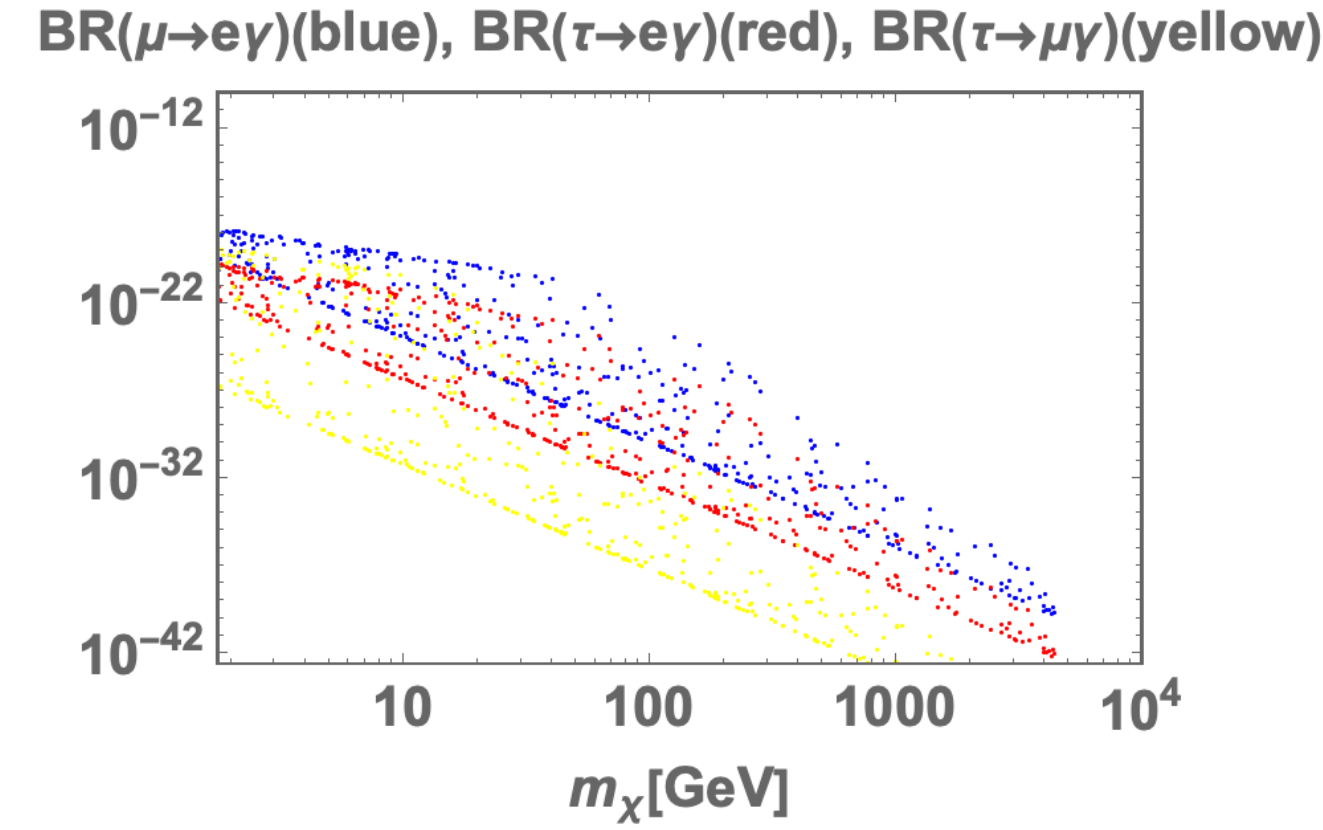} \quad
\includegraphics[width=50.0mm]{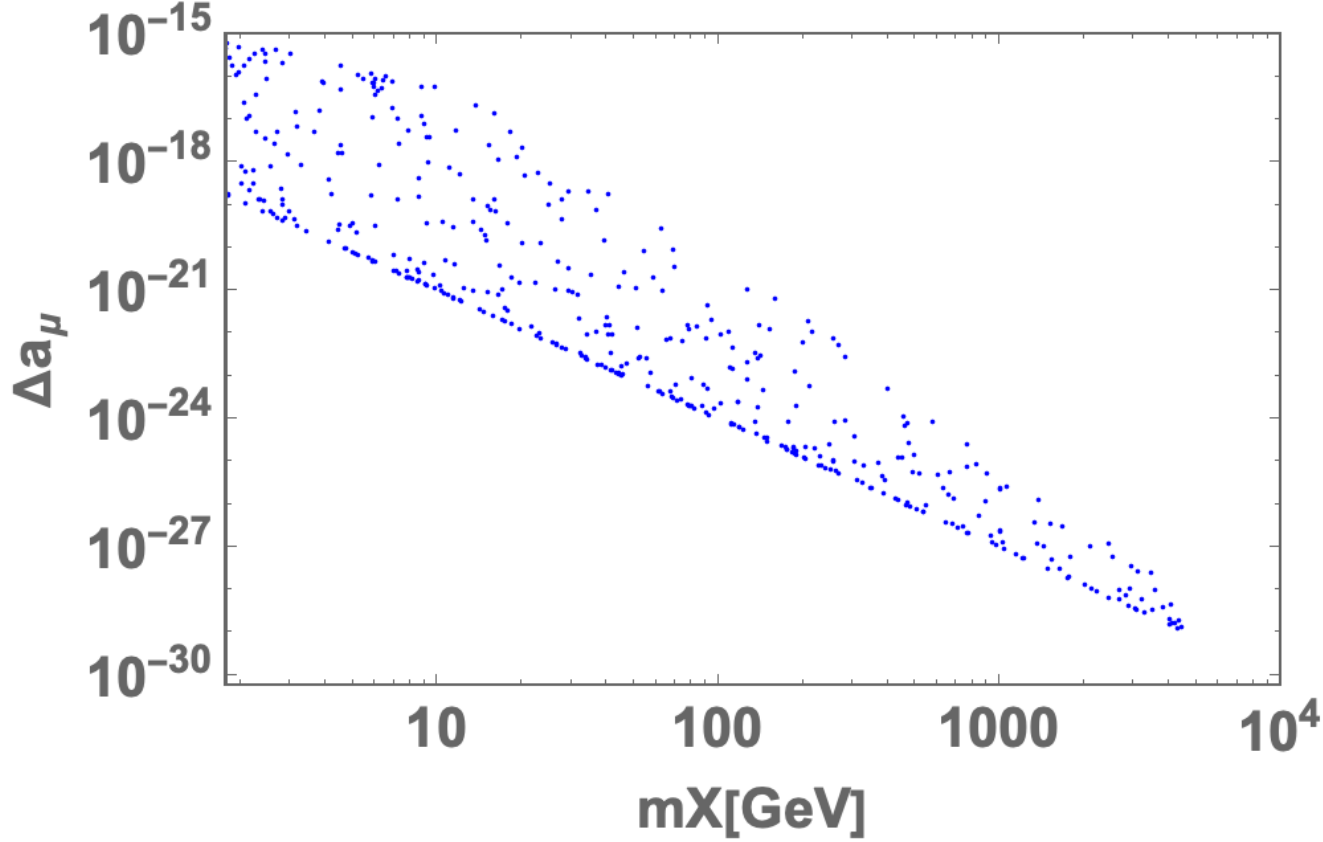} 
\caption{Allowed regions for LFVs (left) and muon $g-2$ (right) in terms of the DM mass. In the left figure,  the blue, red, yellow points correspond  allowed region of BR($\mu\to e \gamma$), BR($\tau\to e \gamma$), and BR($\tau\to e \gamma$), respectively.}
  \label{fig:lfvs_nh}
\end{center}\end{figure}
In Fig.~\ref{fig:lfvs_nh}, we display the LFVs (left) and muon $g-2$ (right) in terms of the DM mass in GeV unit where these points satisfy the relic density of DM at 2$\sigma$; $0.1196\pm2\times0.0031$.
In left figure, the blue, red, yellow points correspond  allowed region of BR($\mu\to e \gamma$), BR($\tau\to e \gamma$), and BR($\tau\to e \gamma$), respectively. As can be seen in these figures, all the LFVs are far from the experimental upper bounds that are totally safe, and muon $g-2$ is at most $10^{-15}$ that is also within the current experimental region.

\subsection{IH}
\begin{figure}[tb]
\begin{center}
\includegraphics[width=50.0mm]{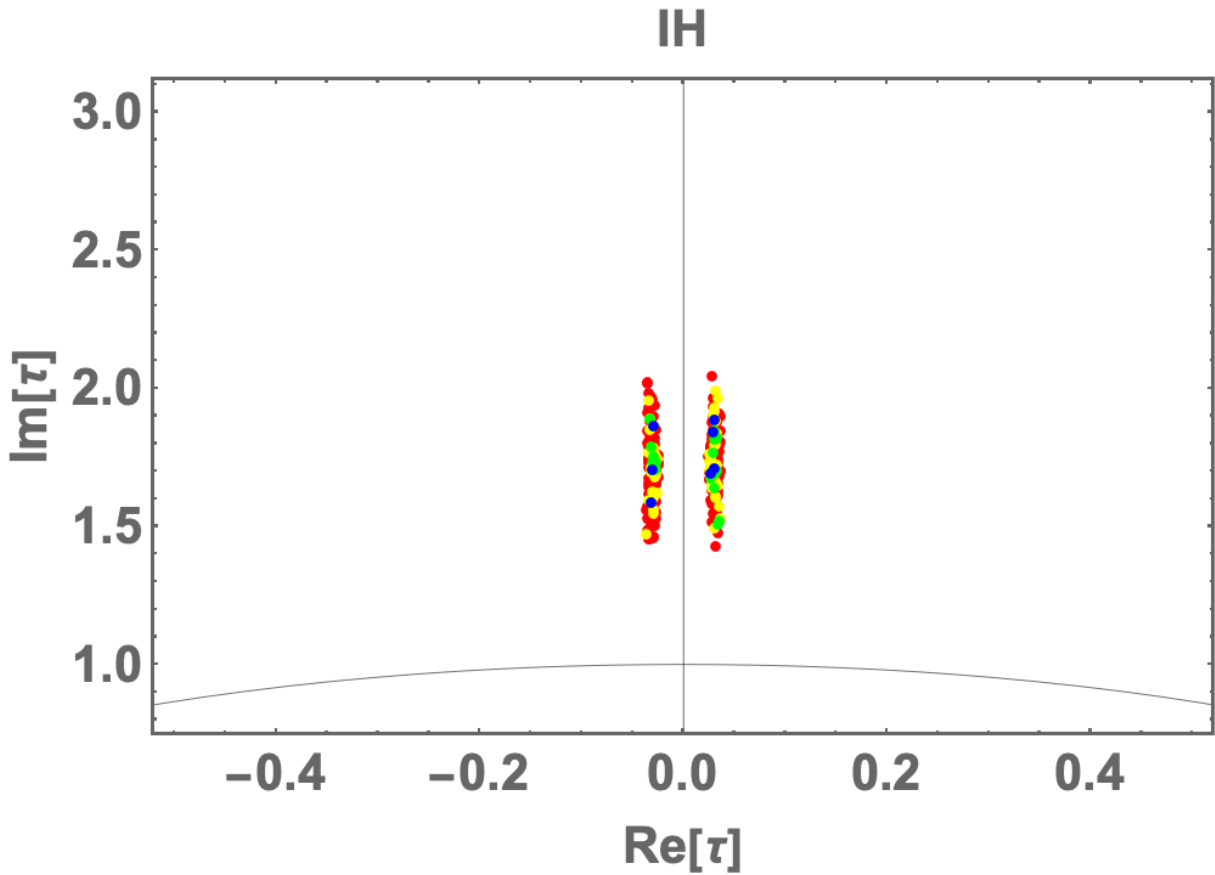} \quad
\caption{Allowed region for Im[$\tau$] in terms of Re[$\tau$] within the fundamental region through the numerical $\chi^2$ analysis.
Here, the red points are within 5$\sigma$, yellow ones within 3$\sigma$, green ones within 2$\sigma$, and, blue ones within 1$\sigma$.}
  \label{fig:tau_ih}
\end{center}\end{figure}

In Fig.~\ref{fig:tau_ih}, we show allowed region of $\tau$,
where the color legends are the same as the ones of Fig.~\ref{fig:tau_nh}.
The allowed region of Re[$\tau$] runs [0.0224-0.0383], on the other hand
the allowed one of Im[$\tau$] is [1.41-1.78].

\begin{figure}[tb]
\begin{center}
\includegraphics[width=50.0mm]{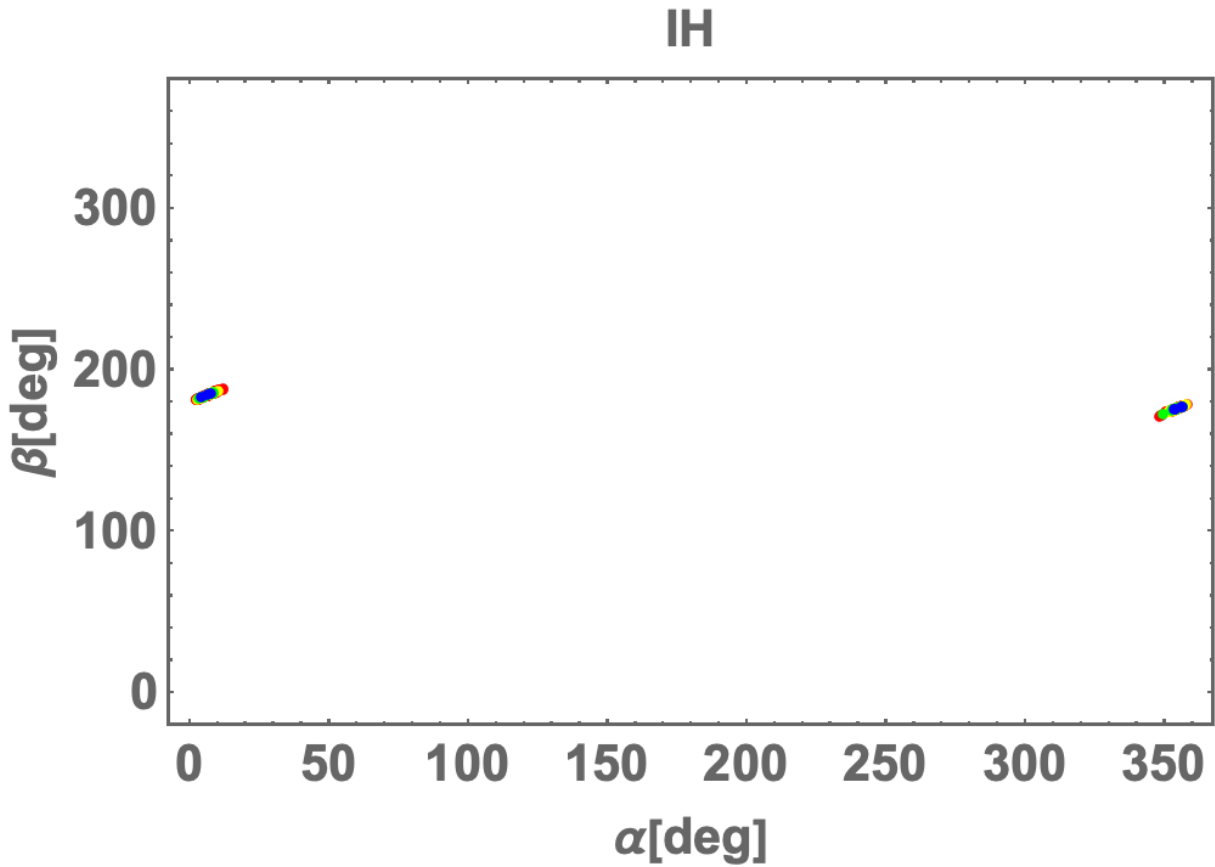} \quad
\includegraphics[width=50.0mm]{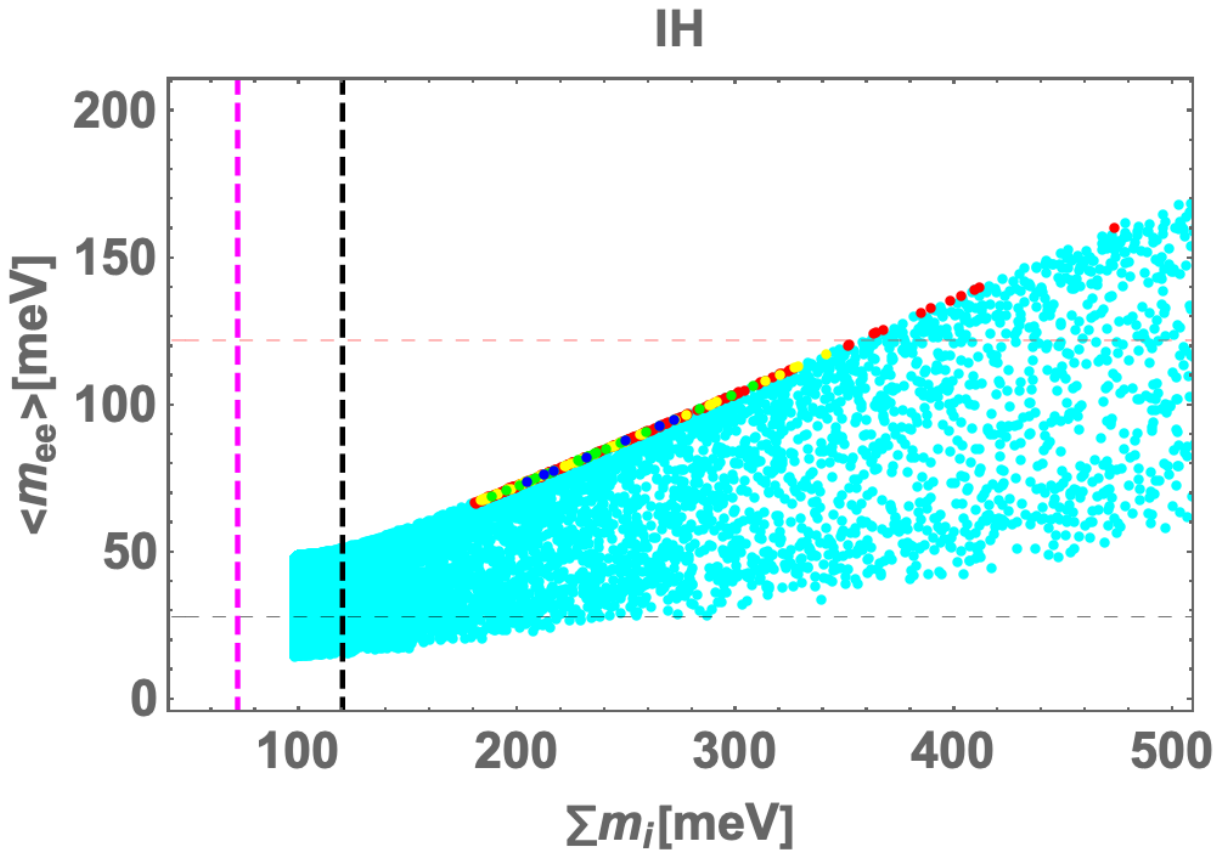} \quad
\includegraphics[width=50.0mm]{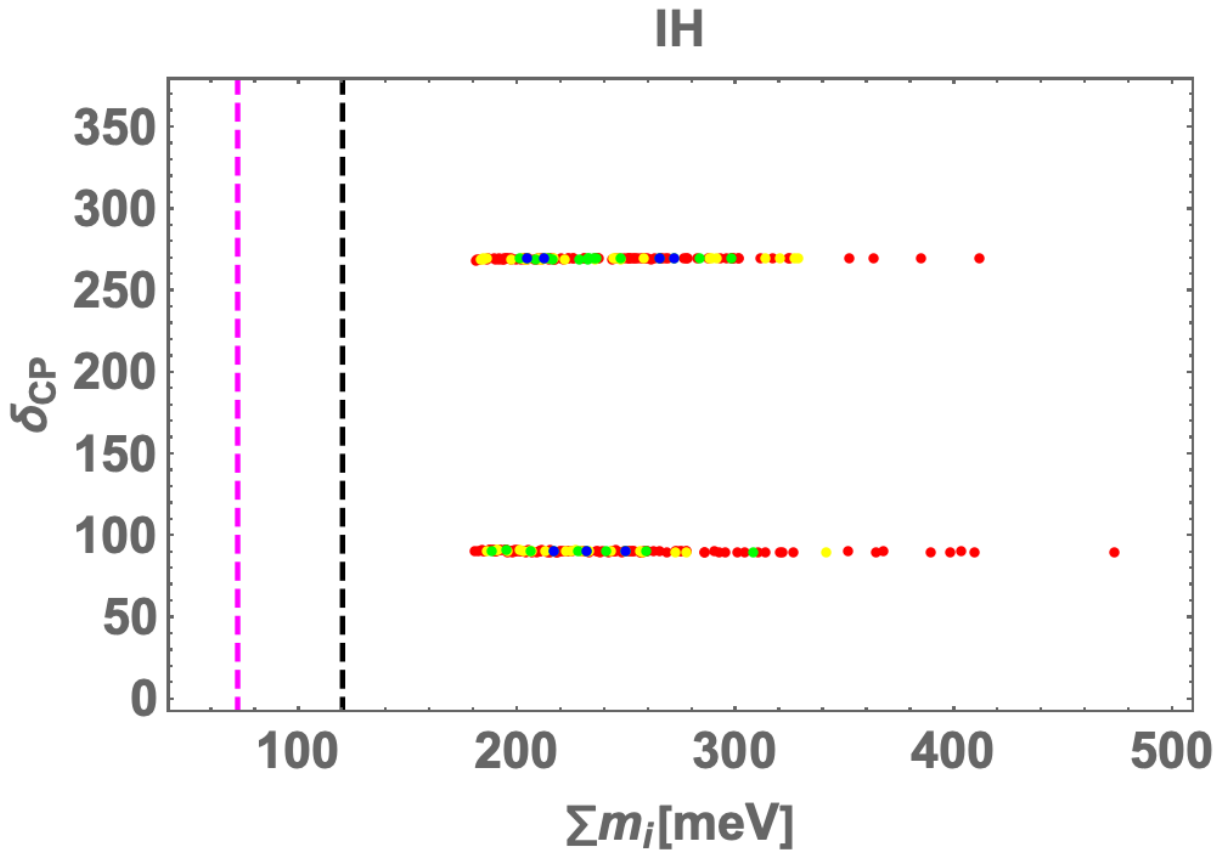} 
\caption{Allowed regions for Majorana phases (left), $\langle m_{ee}\rangle$ (center), and $\delta_{\rm CP}$ (right) in term $\sum m_i$ meV, where all the color legends are the same as the ones of Fig.~\ref{fig:tau_nh}. In the center, the cyan region is experimentally allowed, the magenta and black dotted vertical lines (in the center and right) are respectively the upper bounds of 72 meV (DESI and CMB) and 120 meV (the minimal cosmological model), the horizontal dotted line at 28 meV (in the center) represents the lower upper bound on KamLAND-Zen data. }
  \label{fig:ih2}
\end{center}\end{figure}
In Fig.~\ref{fig:ih2}, we demonstrate the allowed regions for Majorana phases (left), $\langle m_{ee}\rangle$ (center), and $\delta_{\rm CP}$ (right) in term $\sum m_i$ meV, where all the color legends and dotted lines are the same as the ones of Fig.~\ref{fig:nh2}.
We have two localized islands in allowed points of Majorana phases; $\alpha\sim0^\circ$ and $\beta\sim 180^\circ$. 
The sum of neutrino masses is allowed by [180-480] meV and the neutrinoless double beta decay allows [60-150] meV.
The allowed regions of Dirac CP phase have two specific degress; $90^\circ$ and $270^\circ$.

Since IH does not satisfy the upper bound on $\sum m_i\le120$ meV, we will not move on to the discussion of LFVs, muon $g-2$, and our DM candidate.~\footnote{If the vanishing DM mass is allowed, we have the allowed region within the range of $\sum m_i\le$120 meV  for IH. In our paper, we do not consider such a situation.}

\section{Summary and discussion}
We have proposed a two-loop neutrino mass model in which fermionic DM candidate is favored compared to the bosonic one
due to generating the fermionic DM mass at one-loop level.
In order to control our desired  Lagrangian and Higgs potential we have  introduced a $Z_3$ gauging TY non-invertible fusion rule with the assistance of a non-holomorphic modular $A_4$ symmetry. The fusion rule forbids the mass of DM candidate at tree level where its mass is generated at one-loop level where the DM mass term dynamically violates the fusion rule.
After that, the neutrino mass matrix is induced at one-loop level where a remnant $Z_2$ symmetry is remained.
The symmetry assures the stability of our DM candidate.
The non-holomorphic modular $A_4$ symmetry plays a role in forbidding the interactions between the SM particles and $X_R$ and $S_0$ that runs in the $N_R$ loop, in addition to reduction our free parameters that leads to our predictions for lepton sector.
We have performed $\chi^2$ numerical analysis for the lepton masses, mixing angles, and phases, and we have found several localized allowed regions for NH and IH. But IH does not satisfy the upper bound on the sum of neutrino masses $\sum m_i\le$120 meV.  Thus, we have concentrated on analysis of the LFVs, muon $g-2$, and our fermionic DM candidate for the case of NH only.
We have found allowed regions of LFVs and muon $g-2$ satisfying the relic density of DM at 2$\sigma$,
and all the maximum values are far below the current experimental bounds which are totally safe.

\begin{acknowledgments}
 HO is supported by Zhongyuan Talent (Talent Recruitment Series) Foreign Experts Project. 
\end{acknowledgments}

\bibliography{ctma4.bib}

\end{document}